\documentclass[aps,prapplied,english,reprint,amsmath,amssymb,noeprint,longbibliography]{revtex4-2}

\usepackage[T1]{fontenc}
\usepackage[utf8]{inputenc}
\usepackage{amssymb}
\usepackage{graphicx}
\usepackage{babel}
\usepackage{xcolor}
\usepackage{soul}
\usepackage[colorlinks=true,allcolors=black]{hyperref}
\usepackage{siunitx}

\begin{document}

\title
{Two-tone spectroscopy of high-frequency quantum circuits with a Josephson emitter}

\author{A. Peugeot$^{1}$, H. Riechert$^{1}$, S. Annabi$^{1}$, L. Balembois$^{2}$, M. Villiers$^{3}$, E. Flurin$^{2}$, J. Griesmar$^{1}$, E. Arrighi$^{1}$, J.-D. Pillet$^{1*}$, L. Bretheau$^{1}$}
\selectlanguage{english}

\altaffiliation{These authors supervised equally this work.
\newline
jean-damien.pillet@polytechnique.edu
\newline
landry.bretheau@polytechnique.edu}

\affiliation{$^{1}$
Laboratoire de Physique de la Mati\`ere condens\'ee, CNRS, Ecole Polytechnique, Institut Polytechnique de Paris, 91120 Palaiseau, France}
\affiliation{$^{2}$
Quantronics group, Universit\'e Paris-Saclay, CEA, CNRS, SPEC, 91191 Gif-sur-Yvette, France}
\affiliation{$^{3}$
Laboratoire de Physique de l'Ecole normale sup\'erieure, Centre Automatique et Syst\`emes, Mines Paris, Inria, ENS-PSL, Universit\'e PSL, CNRS, Sorbonne Universit\'e, Paris, France}

\begin{abstract}
We perform two-tone spectroscopy on quantum circuits, where high-frequency radiation is generated by a voltage-biased superconductor-normal-superconductor Josephson junction and detection is carried out by an ancillary microwave resonator. We implement this protocol on two different systems, a transmon qubit and a $\lambda/4$ resonator. We demonstrate that this two-tone Josephson spectroscopy operates well into the millimeter-wave band, reaching frequencies larger than 80 GHz, and is well suited for probing highly coherent quantum systems.
\end{abstract}

\maketitle

\section{Introduction}

Circuit quantum electrodynamics (cQED) is based on microwave measurements and control of superconducting circuits~\cite{Blais_Cavity_2004,wallraff_strong_2004,Schoelkopf2008,Blais2020}. In practice however, the accessible frequencies are typically smaller than 30~GHz, due to the limited range of commercially available instruments and components and the difficulty to route high-frequency signals in a cryostat. Increasing this upper limit could help to probe quantum systems in hybrid architectures~\cite{Clerk2020} as well as elucidate decoherence mechanisms caused by the environment at high frequency~\cite{Huang2021}. It would also open alternative routes towards the development of high-frequency qubits that can operate at higher temperatures~\cite{Anferov2024}. In order to take a step forward in this direction, one can exploit the ac Josephson effect~\cite{Josephson1964,Likharev1986} to generate high-frequency radiation directly at the cold-stage level. When biased at voltage $V_J$, a Josephson junction indeed radiates photons at frequency $f_J = 2eV_J / h$, thereby working as an on-chip tunable-microwave source~\cite{Yanson1965,Langenberg1966,Dayem1966,Varmazis1977,Crozat1978,Hansen1979,Deacon2017,Cassidy2017,Yan2021}. This phenomenon enables broadband spectroscopy that extends from a few tens of MHz up to the THz regime. Its principle relies on power conservation and is quantumly described by the theory of
dynamical Coulomb blockade~\cite{Devoret1990,Ingold1992}. 
Each time a photon of energy $h f_J$ is absorbed and dissipated in the junction's electromagnetic environment, a Cooper pair of energy $2eV_J$ inelastically flows across the junction. Consequently, when $f_J$ matches a resonance of the target system, one can detect a current $I_J = 2e \Gamma_J$ that is directly proportional to the photon emission rate $\Gamma_J$.

Absorption Josephson spectroscopy has been employed for characterizing macroscopic systems, such as ensembles of cobalt atoms ~\cite{Silver1967} or microwave resonators~\cite{Pedersen1972,Pedersen1976,Edstam1994,Holst1994,Jack_nanoscale_2015,Griesmar2021}, as well as elementary quantum objects, including Cooper-pair boxes~\cite{Lindell2003,Leppakangas2006,Billangeon2007,Basset2010} and Andreev bound states~\cite{Bretheau2013,Bretheau2013a,Bretheau2013b,VanWoerkom2017}.
This spectroscopy technique is, however, challenging as it is sensitive to the whole electromagnetic environment of the junction and relies on low-frequency measurements of small dc currents. Even more crucially, this approach is not suited to probe highly coherent quantum systems like superconducting qubits, as its detection sensitivity decreases dramatically for long coherence time. Indeed, the dissipation rate of $\sim 1$ photon per coherence time interval induces an upper bound for $\Gamma_J$. A typical $T_1 \sim \SI{100}{\us}$ would thus amount in detecting a current of a few fA, $4$ orders of magnitude below the resolution of the best power-absorption Josephson spectrometer~\cite{Griesmar2021}. 

In this work, we develop another paradigm for Josephson spectroscopy, inspired by cQED, where the detection is realized using an ancillary system (see Fig.~\ref{fig1}a). First, we utilize a superconductor-normal-superconductor (SNS) Josephson junction as an on-chip microwave emitter (red). We couple such an SNS emitter to a quantum object of interest with frequency $f_q$ (green). This system is in turn dispersively coupled to a microwave cavity that acts as a detector (blue) resonating at a readily accessible frequency $f_r$. When driven at resonance $f_J = f_q$, the quantum object gets excited with $n_q$ photons, which induces a frequency shift $\chi n_q$ of the cavity that can be detected using sensitive and fast radiofrequency techniques. Following this approach, our goal is to perform two-tone Josephson spectroscopy at high frequency. As a proof of concept, we implement this strategy to probe two coherent though very different quantum objects, a superconducting transmon qubit that resonates at 6~GHz, and electromagnetic modes of a $\lambda / 4$ resonator up to 83~GHz.

\section{SNS Josephson junction as a microwave emitter}

\begin{figure}
\includegraphics[width=0.9\columnwidth]{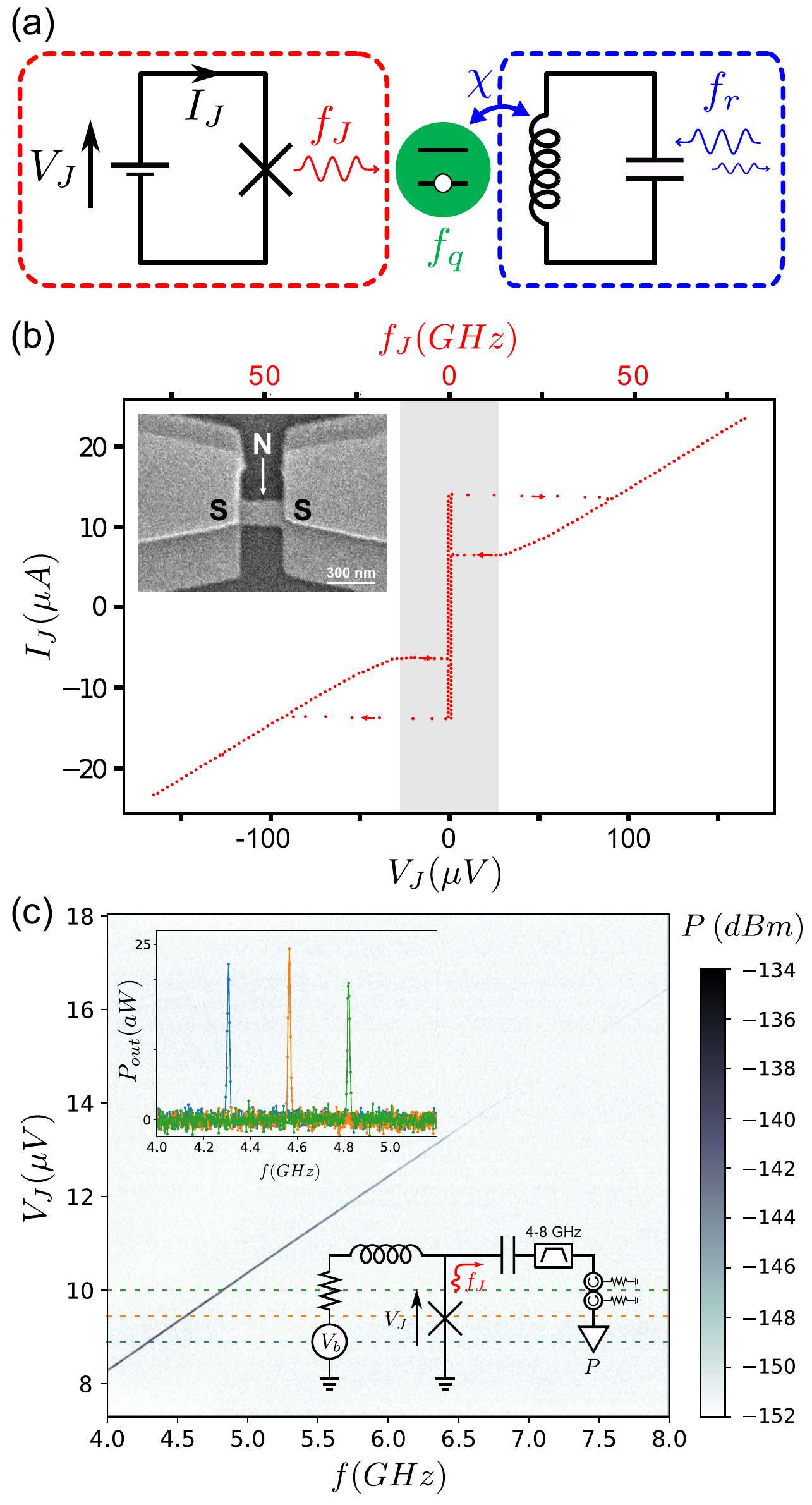}
\caption{\label{fig1}
(a) Principle of two-tone Josephson spectroscopy.
A Josephson emitter (red) emits photons at frequency $f_J=2eV_J/h$ onto the system of interest (green) that resonates at frequency $f_q$. A microwave resonator (blue) with frequency $f_r$ is dispersively coupled to the system of interest and serves as a detector. When the system of interest is excited, the resonator's frequency shifts, 
%to $f_r-\chi$, 
which is detected by a probe tone.
(b) Current-voltage characteristic $I_J(V_J)$ of an SNS Josephson junction 
measured at 7~mK. The gray region $|V_J| \le \SI{23}{\uV}$ is not accessible as 
the current falls below the retrapping current and the junction jumps to the zero-voltage state. Inset: scanning electron microscope image of the device. The normal part is a Ti(5~nm)-Au(15~nm) bilayer and the superconducting part is a 55~nm-thick layer of Nb.
(c) Josephson power $P$ as a function of both frequency $f$ and voltage $V_J$ across the emitter. The Josephson radiation is here emitted by an Al(100~nm)-Hf(7~nm)-Al(100~nm) junction with $I_c\approx \SI{400}{nA}$ and measured using a spectrum analyzer by integrating the power in a 2~MHz bandwidth around $f$.
Cuts corresponding to the blue ($V_J=\SI{8.9}{\uV}$), orange ($V_J=\SI{9.45}{\uV}$) and green ($V_J=\SI{10}{\uV}$) dashed lines are shown in the top left inset. Bottom right inset: simplified version of the setup, where a $4-8$~GHz band-pass filter protects the junction from the $1.5$~K noise emitted by the HEMT.
}
\end{figure}

Our Josephson emitter consists of two superconducting electrodes connected by a normal metallic wire, forming an SNS Josephson junction (see micrograph in Fig.~\ref{fig1}b and details of fabrication in Appendix~\ref{fabrication}). We conduct measurements either on Nb-Au-Nb or Al-Hf-Al structures. In the latter case, although the Hf wire might actually become superconducting at the base temperature of our experiments~\cite{Safonova2024}, it essentially behaves like a normal metal  above its critical current at finite voltage.

We initially characterize our devices through dc electrical measurements and extract their current-voltage characteristic $I_J(V_J)$ (see Fig.~\ref{fig1}b and Appendix \ref{characterization}). Depending on the thickness, width, length and composition of the normal wire connecting the two superconductors, the critical current $I_c$ can vary from a few hundreds nA to a few tens of \unit{\uA}. When biased above $I_c$, the junction switches and acquires a finite normal resistance $R_n$, ranging from about 10 to a few hundreds of Ohms depending on the devices. Due to Joule heating, retrapping occurs at a current significantly smaller than $I_c$. The thermalization is mostly ensured by the normal metal wings, which act as thermal baths and maintain the emitters temperature below a few hundreds of mK. In most of our designs, the $I_c R_n$ product is around \SI{100}{\uV}, which is consistent with results of other experiments with similar geometries~\cite{Courtois2008,Basset2019}. This suggests that our SNS junctions possess a minigap $2E_g\sim\SI{25}{GHz}$ and are likely to exhibit additional dissipation beyond this frequency~\cite{Dassonneville2018}.

The operation of SNS microwave sources relies on the ac Josephson effect, as they carry, when voltage-biased, an oscillating current
at the Josephson frequency $f_J=2 e V_J / h$. The voltage-to-frequency conversion ratio is thus given by the inverse of the magnetic flux quantum $\phi_0^{-1}=2e/h \approx \SI{0.5}{GHz/\uV}$. Consequently, Josephson junctions can emit microwave photons ranging from a few GHz up to a few hundreds of GHz. This upper bound is set by the value of the superconducting gap, which can reach $\sim 180$~GHz, 1.4~THz and 1.8~THz for Al, Nb and NbN respectively, the three most commonly used materials in superconducting circuits. Beyond this limit, the amplitude of the microwave signal is expected to decrease slowly~\cite{Langenberg1974,Poulsen1974}. In the $I_J(V_J)$ characteristic of Fig.~\ref{fig1}b, the ac Josephson current is averaged out and thus invisible. To detect it, we measure the power spectral density emitted by the junction in the 4-8 GHz band by routing the signal with a bias tee to a cryogenic HEMT amplifier anchored at 4K. In the power spectra presented in Fig. \ref{fig1}c, we observe an emission peak whose frequency coincides with $\phi_0^{-1}V_J$, demonstrating that this signal is indeed Josephson emission.

The peak has a Gaussian shape with a standard deviation $\sigma \approx 5$~MHz,
which we attribute to a 10~nV RMS noise on the bias voltage across the junction of both thermal and technical origin~\cite{peugeot_quantum_2020}. Such a thin linewidth, obtained thanks to significant filtering and thermalization of the bias lines (details in Appendix~\ref{setup}), is comparable to the best performances achieved so far~\cite{Hofheinz2011,Rolland2019,Peugeot2021,albert_microwave_2024}. Additionally, we observe a frequency dependence of the signal amplitude within the $4-8$~GHz bandwidth of our detection chain,  with a large decrease of the signal around 7.3~GHz, probably due to a parasitic resonance on the junction chip or in our setup. This illustrates how challenging it is to efficiently route the signal emitted by the junction in the desired direction and to determine it quantitatively.
By integrating the signal from Fig.~\ref{fig1} over the full linewidth and comparing it to the noise floor added by the HEMT amplifier, we find an emission power of $\sim -130$~dBm. This microwave power corresponds to the emission of about $10^7$ photons per second, \emph{i.e.} of the order of magnitude of the read-out pulses used in circuit-QED experiments. Note that the power emitted by the SNS junction strongly depends on the environment it is embedded into and might change for each experiment. At larger voltage, we observe an additional broadband emission due to shot noise (see Appendix~\ref{characterization}), whose impact is negligible in the following experiments.

These dc and ac measurements therefore show that SNS junctions function properly as microwave sources, at least in the $4-8$~GHz range. Interestingly, SNS microwave sources offer several advantages compared to more standard tunnel Josephson junctions. First, their $I_J(V_J)$ characteristic is mostly linear, with $R_n$ remaining relatively small. Their impedance thus weakly changes with $V_J$ and, as such, does not significantly affect other components to which it is coupled. This aspect is crucial for applications like two-tone spectroscopy, where a sensitive microwave cavity interacts with the junction (see Appendix~\ref{tunnel}). Second, one can mention the relative ease of their fabrication process compared to tunnel junctions when using niobium as a superconductor~\cite{Bumble2009,Tolpygo2015,Grimm2017}. Finally, the low backaction from the environment on their $I_J(V_J)$ characteristics make them simpler to use as their resistance does not significantly vary depending on the objects to which they are coupled. In the following, we use these on-chip SNS microwave emitters to perform two-tone spectroscopy of quantum circuits.

\section{Two-tone Josephson spectroscopy of a transmon qubit}
\label{sectiontransmon}

\begin{figure}
\includegraphics[width=0.95\columnwidth]{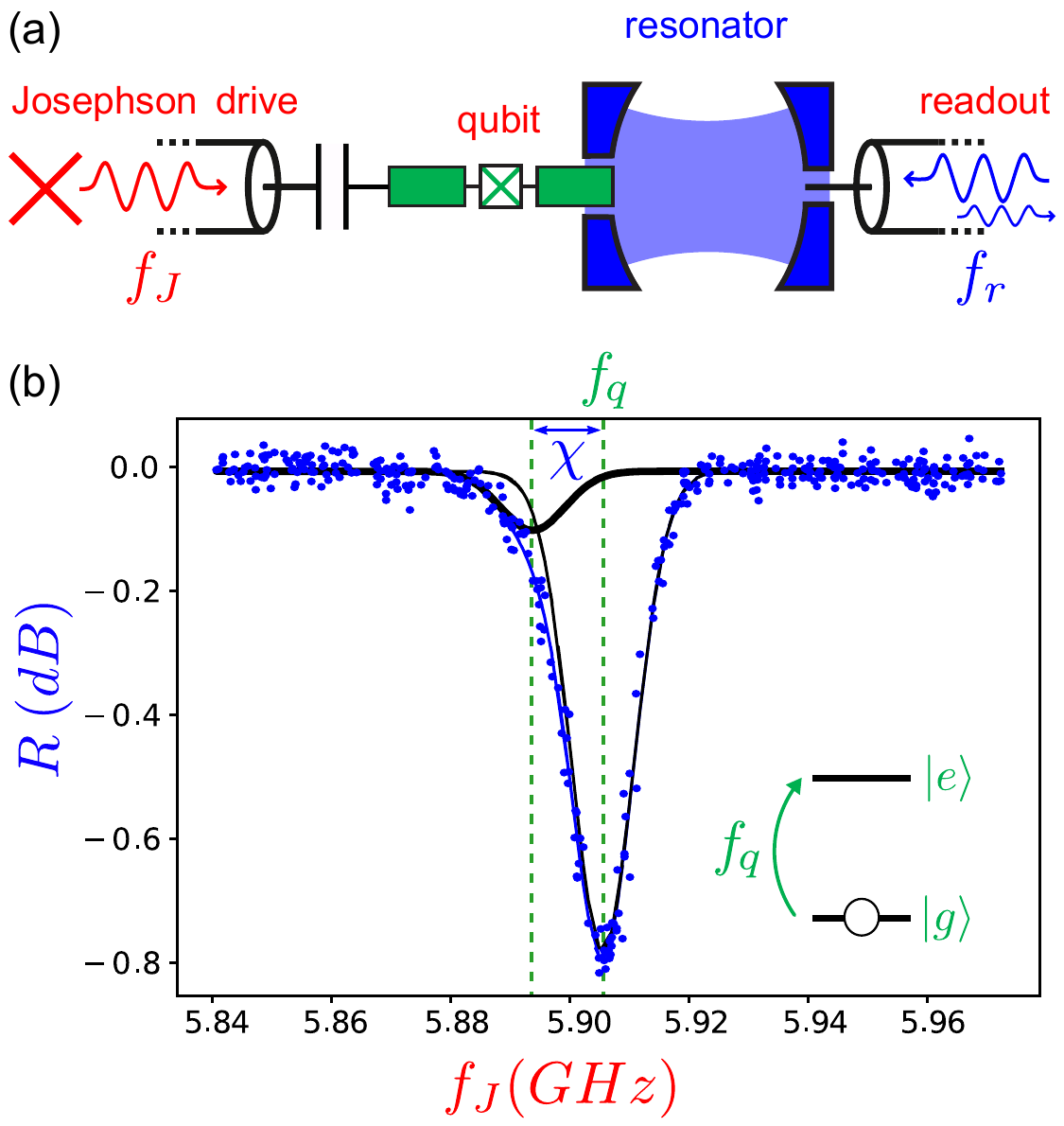}
\caption{\label{fig2}
Two-tone Josephson spectroscopy of a transmon qubit.
(a) Schematic of the experimental setup. An Al-Hf-Al SNS Josephson emitter (red), mounted in its own sample holder, emits photons at frequency $f_J$. These photons are routed via an SMA cable to another sample holder towards the drive line of a transmon qubit (green), which is dispersively coupled to a microwave resonator (blue). By probing the resonator close to its resonant frequency $f_r$, one can detect the excitation of the qubit. Though not represented, a $4-8$~GHz bandpass filter is also present in the setup. 
(b) Reflection coefficient $R$, normalized by its value in absence of the Josephson drive, as a function of the emitted Josephson frequency $f_J$. Experimental data are shown as blue dots and fits as solid lines. The latter corresponds to the sum of two Gaussians of standard deviation $\sigma$ centered at $f_q$ and $f_q-\chi$, with a relative weight $p_1$ given by the finite population of the Fock state $|n=1 \rangle$ in the resonator. Fit parameters: $f_q = \SI{5.9055}{GHz} \pm \SI{0.1}{MHz}$, $\chi = 11.8 \pm \SI{0.8}{MHz}$, $p_e = 0.13 \pm 0.02$, $\sigma = 5.4\pm \SI{0.1}{MHz}$.
}
\end{figure}

As a first proof of concept, we employ two-tone Josephson spectroscopy on a 
well-known quantum system: a transmon qubit~\cite{koch_charge-insensitive_2007}. While its frequency $f_q$ may be relatively low, it allows us to validate our technique by comparing it to measurements obtained using conventional microwave instrumentation. The transmon is capacitively coupled to a microwave resonator of frequency $f_r$ for dispersive readout (see Fig.~\ref{fig2}a). The qubit and cavity being off-resonantly coupled with a large enough detuning, they can be described by the Hamiltonian $h f_r a^\dagger a + h f_q |e\rangle \langle e| - h \chi a^\dagger a |e\rangle \langle e| $, where $a^\dagger$ and $a$ are the ladder operators of the cavity, $|g\rangle$ and $|e\rangle$ the ground and excited state of the qubit, and $\chi$ is the dispersive frequency shift. The cavity frequency is thus conditioned on the state of the qubit, which provides the basis for performing its two-tone spectroscopy.
Details on the design and fabrication of this circuit, which was initially created for microwave photon counting are given in Refs.~\cite{Balembois2023,Balembois2024}.

We first characterize the transmon circuit using standard microwave instruments and thus access to its characteristic frequencies and coherence times (see Appendix~\ref{RF}). We then turn to two-tone Josephson spectroscopy using our on-chip SNS microwave emitter ($I_c\approx\SI{400}{nA}$, see Fig. \ref{fig5} in Appendix \ref{characterization}). For this experiment, the emitter and the transmon are separated on two independent chips, each mounted in its own sample-holder, connected to each other by an SMA cable (see Fig.~\ref{fig2}a). We emit Josephson radiation onto the transmon while probing the resonator via reflectometry close to its resonant frequency $f_r=\SI{8.177}{GHz}$. As we vary $f_J$, we observe a clear suppression of the reflection coefficient $R$ at the qubit frequency $f_q=\SI{5.9055}{GHz}$, demonstrating that it gets excited by the SNS microwave source (see Fig.~\ref{fig2}b). 

The transition linewidth with $\sigma = \SI{5.4}{MHz}$ is much larger than the intrinsic qubit decay rate. This width is set by the voltage noise across the SNS emitter and is similar to the one measured in Fig.~\ref{fig1}c. On top of that, the resonant dip displays an asymmetry, as a second transition can be excited at frequency $f_J = f_q - \chi$, with $\chi = \SI{11.8}{MHz}$.  The qubit frequency indeed depends on the number of photons $n$ in the cavity due to the dispersive coupling. As a consequence, we observe a transition between states $|g,1\rangle$ and $|e,1 \rangle$, owing to a finite population $p_1 \approx 0.1$ of the Fock state $|n=1 \rangle$. We therefore demonstrate that our on-chip SNS microwave emitter can be used to detect an elementary quantum system, a transmon qubit that resonates around 6~GHz, using two-tone Josephson spectroscopy.

\section{Two-tone Josephson spectroscopy of high-frequency modes}
\label{sectionlambda4}

\begin{figure*}
\includegraphics[width=0.95\textwidth]{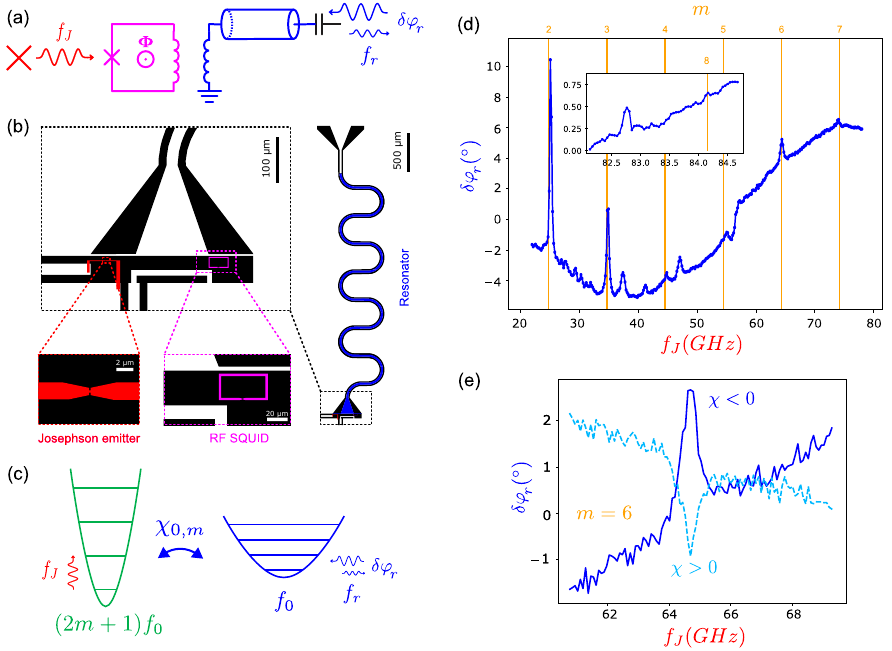}
\caption{\label{fig3}
Two-tone Josephson spectroscopy of high-frequency modes.
(a) Schematic of the circuit, 
and (b) illustration of the device layout (with niobium in white and silicon in black). The $\lambda/4$ resonator (blue) is inductively coupled at its end, both to
a superconducting loop closed by an SNS junction (rf SQUID in pink) and to an SNS Josephson emitter (red). Both Josephson junctions are Nb-Au-Nb structures. A flux line, visible at the bottom, is used to adjust the magnetic flux $\Phi$ in the rf SQUID loop.
(c) Diagram illustrating the principle of two-tone spectroscopy based on the cross-Kerr effect between electromagnetic modes within the same resonator. The fundamental mode $f_0$ is continuously monitored with a probe tone $f_r$, while the $m$-index mode is irradiated at frequency $f_J$. At resonance $f_J = f_m$, mode $m$ gets excited with $n_m$ photons. Consequently, the fundamental mode frequency shifts to $f_0 - \chi_{0,m}n_m$ and a phase shift $\delta\varphi_r$ of the reflected signal is observed.
(d) Phase shift $\delta\varphi_r$ of the probe tone as a function of $f_J$. Orange vertical lines serve as markers for the expected frequencies $f_m$ of the resonator's harmonics. Close to each line, a resonance corresponding to an electromagnetic mode of the resonator is observed. Inset: detailed view around a resonance close to 83~GHz, interpreted as the $m=8$ mode of the resonator. The signal-to-noise ratio was here optimized by carefully selecting the probe tone's frequency $f_r$ and the magnetic flux $\Phi$.
(e) Detailed view of the resonance for mode $m=6$, measured at $\Phi=0$ (solid line) and $\phi_0/2$ (dashed line). The observed sign change of the dispersive coupling $\chi_{0,6}$ confirms that this resonance corresponds to a mode of the resonator.}
\end{figure*}

To further test this alternative detection scheme, we now probe the electromagnetic modes of a $\lambda/4$ resonator. The advantage of such a system is that it has 
a large number of regularly spaced modes, which is well suited to explore two-tone Josephson spectroscopy at high frequency and over a large range. The resonator, shown in Fig.~\ref{fig3}a-b, is made out of niobium and resonates at its harmonic frequencies $f_m$, where $m \in \mathbb N$ is the mode index. It is inductively coupled to an rf SQUID, which consists of an SNS Josephson junction enclosed in a superconducting loop threaded by a magnetic flux $\Phi$. This introduces some nonlinearity into the circuit, which results in cross-Kerr interaction between the resonant modes. The system can thus be described by the Hamiltonian $\sum_m h f_m a^\dagger _m a_m + \sum_{m,m'} h \chi_{m,m'} a^\dagger _m a_m a^\dagger _{m'} a_{m'}$, where $\chi_{m,m'}$ are the cross-Kerr frequencies~\cite{wustmann_nondegenerate_2017}. Therefore, when mode $m$ gets excited with 1 photon, the fundamental mode's frequency $f_0$ is shifted by $\chi_{0,m}$. This provides the basis to perform a two-tone Josephson spectroscopy of the high-frequency modes, while the fundamental mode of the resonator is used as a detector (see Fig.~\ref{fig3}c).

As before, we use a voltage-biased SNS Josephson junction as a microwave emitter. It is positioned on the same chip as the $\lambda / 4$ resonator, with which it is inductively coupled. We emit Josephson radiation while probing the fundamental mode of the resonator at frequency $f_r \approx f_0$. Fig.~\ref{fig3}d shows the phase shift $\delta \varphi_r$ of the reflected signal as a function of the Josephson frequency $f_J$. The spectrum exhibits regularly spaced resonances at frequencies $f_J\approx f_m=(2m+1)f_0$, with a mean relative deviation of about $1~\%$ between measurement and theory. Note that no fitting parameter is used as $f_0 = \SI{4.9572}{GHz}$ is extracted from one-tone spectroscopy measurement of the cavity (see Appendix \ref{Complementary}). These resonances thus correspond to the higher harmonic modes of the $\lambda / 4$ resonator. Other resonances appear in our signal, but are not clearly identified at this stage. We believe they might be geometric resonances of the circuit that would couple to the resonator. Additionally, a slowly varying background is noticeable. It is attributed to a flux change within the loop terminating the resonator, induced by the dc current carried by the Josephson emitter. 

The measured linewidth of the different transitions, which is roughly constant and about $\sim 500$~MHz, is much larger than the intrinsic decay rate of the modes that we estimate to be $\sim 350$~kHz based on the measured quality factor $Q=14000$ of the fundamental mode. It is again limited by the Josephson emission linewidth, which is here almost 2 orders of magnitude larger than previously observed. This larger linewidth is due to a larger voltage noise across the SNS emitter, owing to the different filtering setup needed for this experiment (see Appendix~\ref{setup} and \ref{Linewidth}). Within this architecture, we are able to detect the modes up to $m=8$, corresponding to $ 83$~GHz. We do not detect modes $m \ge 9$, which we attribute to dissipation in the emitter junction shunting the high-frequency Josephson radiation. In essence, the normal resistance $R_N$ of the junction and its coupling inductance $L \sim \SI{100}{pH}$ induce a cutoff frequency at $R_N/L \sim 15$ GHz above which the signal decreases as $1/(f_J)^2$, until it is too low to be detectable within a reasonable averaging time. Although it limits our sensitivity at high frequencies, this cutoff is less drastic than the $LC$ cutoff of tunnel junction-based spectrometers, where the sensitivity is suppressed as $1/(f_J)^4$ \cite{Bretheau2013}. This cutoff frequency can be increased either by opting for an emitter with higher normal resistance or by reducing the junction's length to increase the size of its minigap. This problem is however unlikely to arise when probing a more nonlinear system, such as Andreev bound states, where the coupling $\chi$ with the resonator should be much larger (see section~\ref{Coupling}).

Going further, we adjust the flux $\Phi$ threading the rf SQUID loop. The primary consequence is the modulation of the fundamental mode's frequency $f_0$, with a measured magnitude of $1.5$~MHz. Consequently, the higher harmonic modes $f_m$ are also expected to modulate with $\Phi$. However in practice, this modulation is not observable in the two-tone Josephson spectrum, as it is much smaller than our resolution, limited by the Josephson emission width. Nonetheless, changing the flux from $\Phi =0$ to $\Phi = \phi_0/2 $ strongly alters the spectrum, as illustrated in Fig.~\ref{fig3}e for resonance $m=6$ around 64~GHz. Instead of a peak, a resonant dip is observed at $\Phi = \phi_0/2 $, demonstrating an inverted frequency shift. This change can be understood as a reversal of the Josephson inductance of the rf SQUID when varying the phase drop across the junction~\cite{wustmann_nondegenerate_2017} (see Appendix \ref{theory}), which further confirms the identification of the harmonic modes of the $\lambda / 4$ resonator.

\section{Conclusion and perspectives}

These measurements demonstrate that SNS Josephson junctions behave as on-chip broadband microwave emitters, capable of performing two-tone spectroscopy on superconducting quantum circuits at unprecedented frequencies. With this innovative
technique, we were able to probe both a transmon qubit resonating around $6$ GHz and the electromagnetic modes of a $\lambda / 4$ resonator up to 83~GHz.
Two-tone Josephson spectroscopy is here well adapted to detect these highly coherent systems, with lifetimes $\sim \SI{10}{\us}$ and $\sim \SI{0.5}{\us}$ respectively. All indications suggest that this technique should operate at higher frequencies, albeit necessitating some adjustments in both the emitter properties and its coupling with the system under study.
Conducting the same type of spectroscopy using commercial components, like frequency doublers, presents significant challenges as it would require routing high-frequency signals from room temperature to the cryostat. Two-tone Josephson spectroscopy thus provides a simple and easy to implement method to characterize high-frequency quantum circuits. Greater flexibility can also be achieved by using a flip-chip approach, in which the Josephson emitter and the system to be probed are placed on two separate chips that are brought close together~\cite{Griesmar2021}.

More fundamentally, two-tone Josephson spectroscopy could be instrumental in detecting fermionic or bosonic excitations in mesoscopic systems, such as Andreev bound states in superconducting quantum dots~\cite{Pillet2010,deacon_tunneling_2010} or topological Weyl band structures in multiterminal Josephson junctions~\cite{Fatemi2021,Peyruchat2021,Peyruchat2024}. It could also prove valuable for studying quasiparticle generation~\cite{Liu2024} and recombination in exotic 2D superconductors~\cite{Fatemi2018,Cao2018,Hao2021,Park2022}. Another promising axis is to explore the quantum nature of Josephson radiation~\cite{Hofheinz2011,Jebari2018,Rolland2019,Peugeot2021,Menard2022,Albert2023,Leppakangas2015,Souquet2016,Leppakangas2018a}. This could lead to the development of Josephson photonics devices, such as quantum-limited amplifiers~\cite{Jebari2018} or single-photon detectors~\cite{Leppakangas2018,Albert2023}, operating at high frequencies. 
More generally, the architectures that we developed could be used to probe the high-frequency environment of standard superconducting qubits and how it induces decoherence on them. Finally, an exciting research direction involves sharpening the emitted Josephson frequency, achievable either by reducing voltage noise via even better filtering and thermalization or by employing clever injection locking techniques~\cite{Cassidy2017,Bengtsson_Nondegenerate_2018,Markovic2019,Danner2021,Yan2021}. This advancement would enable time-domain measurement and control at unprecedented frequencies, thereby opening up a realm for high-frequency cQED.

\begin{acknowledgements}
We first want to emphasize the invaluable help of the late F. Portier and P. Jacques from the Nanoelectronics Group on low-noise electronics. We acknowledge valuable discussions with the Quantronics Group, in particular with C. Urbina for sharing the initial idea of two-tone Josephson spectroscopy. Special thanks to \c{C}. Girit, J.-L. Smirr, and Z. Leghtas on microwave measurements. Gratitude is extended to the SPEC of CEA-Saclay for their help on nanofabrication, and to D. Roux and R. Mohammedi from LPMC for their technical support. L.B. acknowledges support of the European Research Council (ERC) under the European Union's Horizon 2020 research and innovation programme (Grant Agreement No.~947707). J.D.P. acknowledges support of Agence Nationale de la Recherche through Grant No. ANR-20-CE47-0003. This work has been supported by the French ANR-22-PETQ-0003 Grant under the France 2030 plan.

\end{acknowledgements}

\begin{appendix}
\renewcommand{\thesubsection}{APPENDIX \Alph{subsection}:}

\section{\label{fabrication}Sample fabrication}

\begin{figure}
    \centering
    \includegraphics[width=0.95\linewidth]{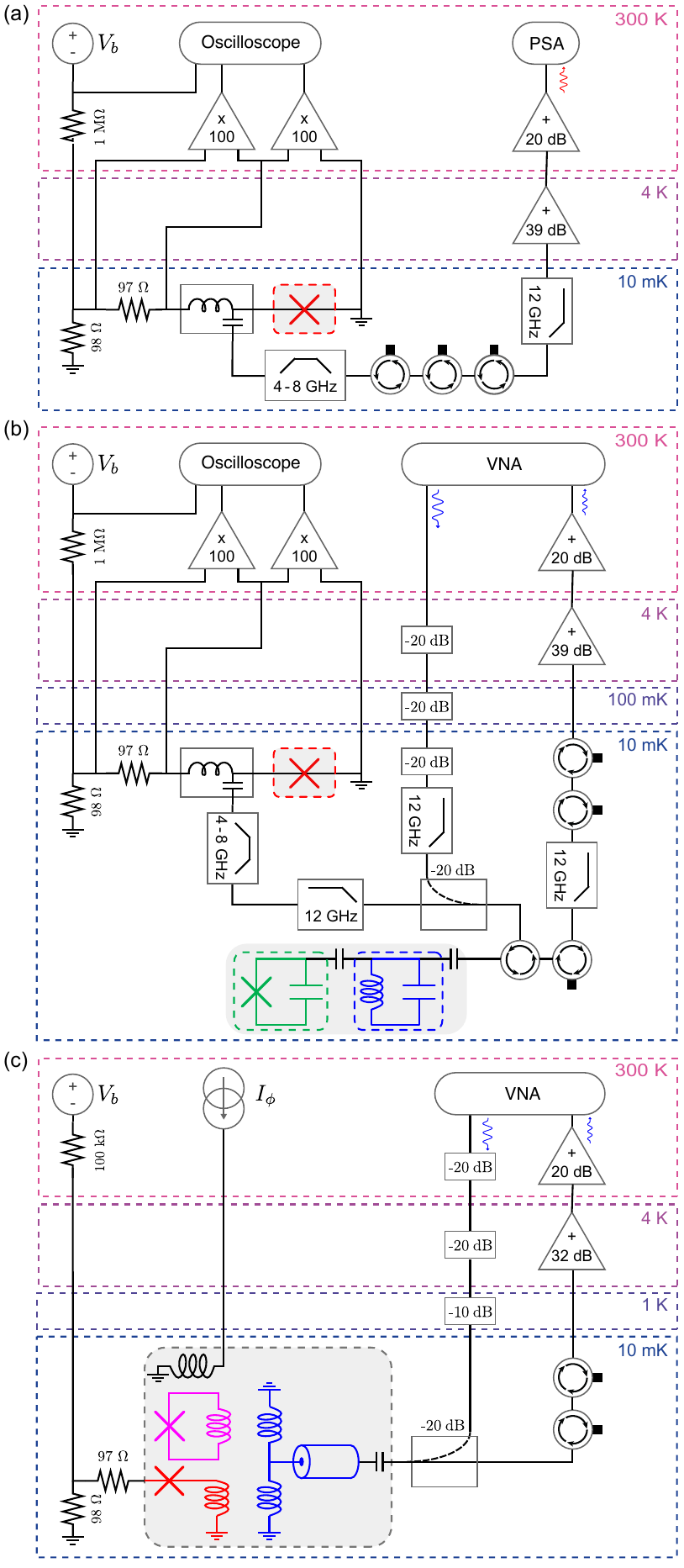}
    \caption{\label{fig4}
    Schematics of the experimental setups for (a) Josephson emission measurements, (b) two-tone Josephson spectroscopy of a transmon qubit and (c) of a $\lambda/4$ resonator.}
\end{figure}

Resonator fabrication begins with sputtering 150~nm of Nb onto an oxidized high-resistivity silicon wafer, which is subsequently diced into $10\times \SI{11}{\square\mm}$ chips. For processing a single chip, a layer of S1813 resist is spun on top of the Nb. Following this, the negative of the resonator pattern is defined using a laser writer and developed in MF319. Subsequently, reactive ion etching is utilized to etch away the Nb through the resist mask. The resist is then removed by immersing the chip in acetone maintained at \SI{60}{\degreeCelsius} for 1~hour, followed by sonication in an ultrasonic bath.

Our SNS junctions are fabricated via metallic deposition on the same chip using a suspended resist mask. For this, a bilayer of MAA EL6 resist is spun at 4000~rpm, followed by a layer of PMMA A6 at a slower speed of 3000~rpm. This results in a bilayer with a thin bottom layer (210~nm) and a very thick top layer (520~nm). The SNS junction mask is defined using e-beam lithography. The bilayer is then developed in a 1:3 MIBK/IPA solution for 1 minute, which provides a suspended mask. The chip is then loaded into a metal evaporator. Before metal deposition, we perform an ion-milling step using an argon plasma to remove the layer of Nb oxide at the chip surface, which could hinder proper electrical contact between the SNS junction and the voltage leads. Initially, 5~nm of Ti is evaporated to serve as a sticking layer for subsequent metals. This is followed by a 15~nm Au deposition. The sample is then tilted by $22^{\circ}$, and 60~nm of Nb is evaporated. Nb is deposited almost everywhere over the gold layer but not at the constriction level, where it is blocked by the mask of PMMA due to the angle, thus preventing from shorting the SNS junction. The pressure inside the loadlock is maintained below a few $10^{-7}$~mbar throughout the process to ensure a clean SN interface. Finally, the resist is lifted off after evaporation in a hot acetone bath.

\section{Measurement setups}
\label{setup}

In Fig. \ref{fig4}, we represent the three implementations of the experiments presented in this paper. 

The setup in Fig. \ref{fig4}a corresponds to one used for the measurement of the Josephson emission, \emph{i.e.}, the experimental data presented in Fig. \ref{fig1}c. The power emitted by the junction is directed, via a bias tee, towards a power spectrum analyzer (PSA) through a ZX60-83LN12 and a room-temperature amplifier as well as a $4-8$~GHz HEMT amplifier from LNF anchored at 4~K. The junction is protected from the noise emitted by the HEMT by means of three $4-8$~GHz cryogenic isolators, a 12~GHz low-pass filter from $K\& L$, and a $4-8$~GHz bandpass filter from Micro-Tronics. This setup also allows for the measurement of the current-voltage characteristic of the Josephson emitter. For this, a voltage is applied with a Yokogawa GS200 source through a $\div 10^4$ voltage divider, and both voltage and current are measured using SP1004 voltage amplifiers from Basel Precision Instrument and a Yokogawa DL350 oscilloscope. The current is determined by measuring the voltage across a resistor in series with the Josephson junction, whose resistance value (\SI{97}{\Omega}) was measured during the run. For all dc lines, we use Thermocoax coaxial cables, RC low-pass filters and $\pi$ filters, both at 10~mK, to protect the junctions from the noise. Detailed references of most components can be found in Ref.~\cite{peugeot_quantum_2020}.

\begin{figure}
    \centering
    \includegraphics[width=0.8\linewidth]{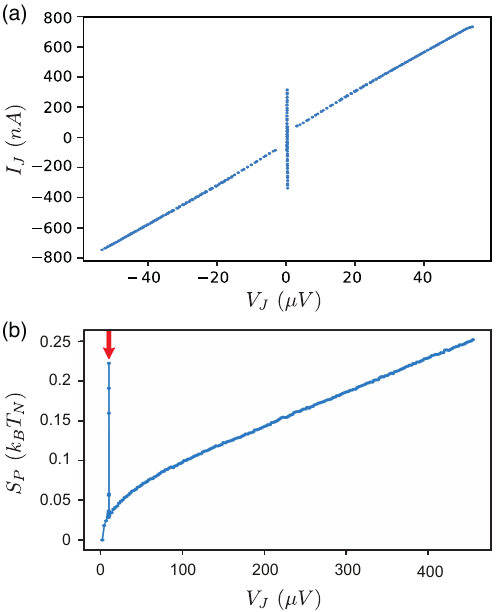}
    \caption{\label{fig5}
    Dc and ac characterization of the Al(100~nm)-Hf(7~nm)-Al(100~nm) Josephson junction.
    (a) Current-voltage characteristic $I_J(V_J)$. The switching current is 385~nA and its normal resistance is \SI{75}{\Omega}.
    (b) Excess noise $S_P$ emitted by the junction at 5~GHz, in units of the HEMT noise background $k_B T_N$. The red arrow indicates the AC Josephson effect peak.}
\end{figure}

In Fig. \ref{fig4}b, we show the setup used for the two-tone Josephson spectroscopy of a transmon qubit (Fig. \ref{fig2}). The dc part of the setup is similar to before, but the rf part is changed significantly. This time the rf port of the bias tee directs the Josephson emission towards the readout port of a transmon qubit through a $4-8$~GHz bandpass filter, a 12~GHz lowpass filter and a circulator. The same port is also used for cavity spectroscopy by sending a microwave signal from a ZNB20 vector network analyzer (VNA) through a $\SI{-60}{dB}$ attenuator, a 12~GHz low-pass filter, and then through the $\SI{-20}{dB}$ port of a directional coupler. The latter allows the VNA signal to be combined with the one emitted by the junction, enabling two-tone spectroscopy of the transmon qubit. Once the microwave signal from the VNA is reflected off the cavity, it is redirected back to the VNA, passing through three 4--8~GHz isolators, a 12~GHz low-pass filter, and a similar amplification chain as before.

The final setup, depicted in Fig. \ref{fig4}c, corresponds to the two-tone Josephson spectroscopy experiment of the $\lambda/4$ resonator shown in Fig. \ref{fig3}. Although not fully represented, the dc part is similar to the previous two setups, with a few minor modifications. Additionally, this setup includes a flux line that allows a current to flow close to the superconducting loop terminating the resonator, enabling us to adjust the magnetic flux in the loop. The spectroscopy of the fundamental mode of the resonator is performed in a standard manner, with a VNA connected to an attenuated input line and an output line with a similar amplification chain as before.

\section{Characterization of SNS junctions}
\label{characterization}

The current-voltage characteristic of the Al-Hf-Al junction, used for the experiments described in Fig.~\ref{fig1} and \ref{fig2}, is shown in Fig.~\ref{fig5}a. Unlike our Nb-Au-Nb junctions, it was not designed for an efficient evacuation of Joule power. Consequently, it exhibits a significantly lower retrapping current compared to its switching current. In order to characterize the Al-Hf-Al junction as a microwave source, we measured the emitted power in a narrow band centered at 5~GHz as a function of the bias voltage (see Fig.~\ref{fig5}b). The narrow peak close to \SI{10}{\uV}, indicated by a red arrow, corresponds to the AC Josephson effect. The continuously increasing background corresponds to incoherent noise, due to shot noise and thermal noise, and remains lower than the AC Josephson emission in the voltage bias range of interest.

\section{Calibration of the Josephson frequency}
\label{Calibration}

\begin{figure}
    \centering
    \includegraphics[width=0.9\linewidth]{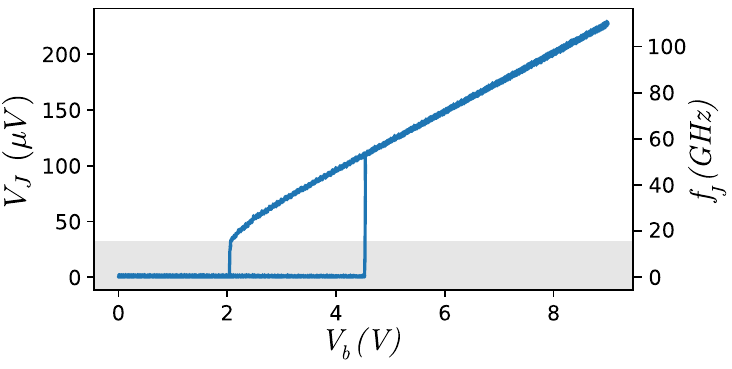}
    \caption{\label{fig6}
    Voltage $V_J$  across the Nb-Au-Nb SNS junction measured as a function of the bias voltage $V_b$. This allows us to calibrate the Josephson frequency $f_J$, which is displayed on the right y axis. The gray region is not accessible as the biasing is unstable due to retrapping of the junction.}
\end{figure}

In order to properly calibrate the frequency $f_J$ of the signal emitted by the SNS junction, it is crucial to accurately know the voltage $V_J$ across its terminals. As we apply a voltage $V_b$ to the bias line of the junction, we simultaneously measure $V_J$ using a low-noise amplifier. As depicted in Fig.~\ref{fig6}, these two voltages are related by the equation $V_J^2 = \epsilon (V_b-V_\text{offset})^2 - V_{0}^2$, which is consistent with the RCSJ model~\cite{mccumber_effect_1968}. Here $\epsilon$ is a constant that depends on the attenuation of our line, $V_\text{offset}$ is the voltage offset between the room-temperature bias source and the sample extracted from the I(V) characteristic, using the supercurrent branch as a reference, and $V_{0}$ is a constant extracted from the fit. This relation allows us to determine the actual value of $V_J$, and thus the Josephson frequency $f_J = 2eV_J / h$, without constantly monitoring it when performing two-tone Josephson spectroscopy.

\section{Standard rf measurements of the transmon qubit}
\label{RF}

\begin{figure}
    \centering
    \includegraphics[width=\linewidth]{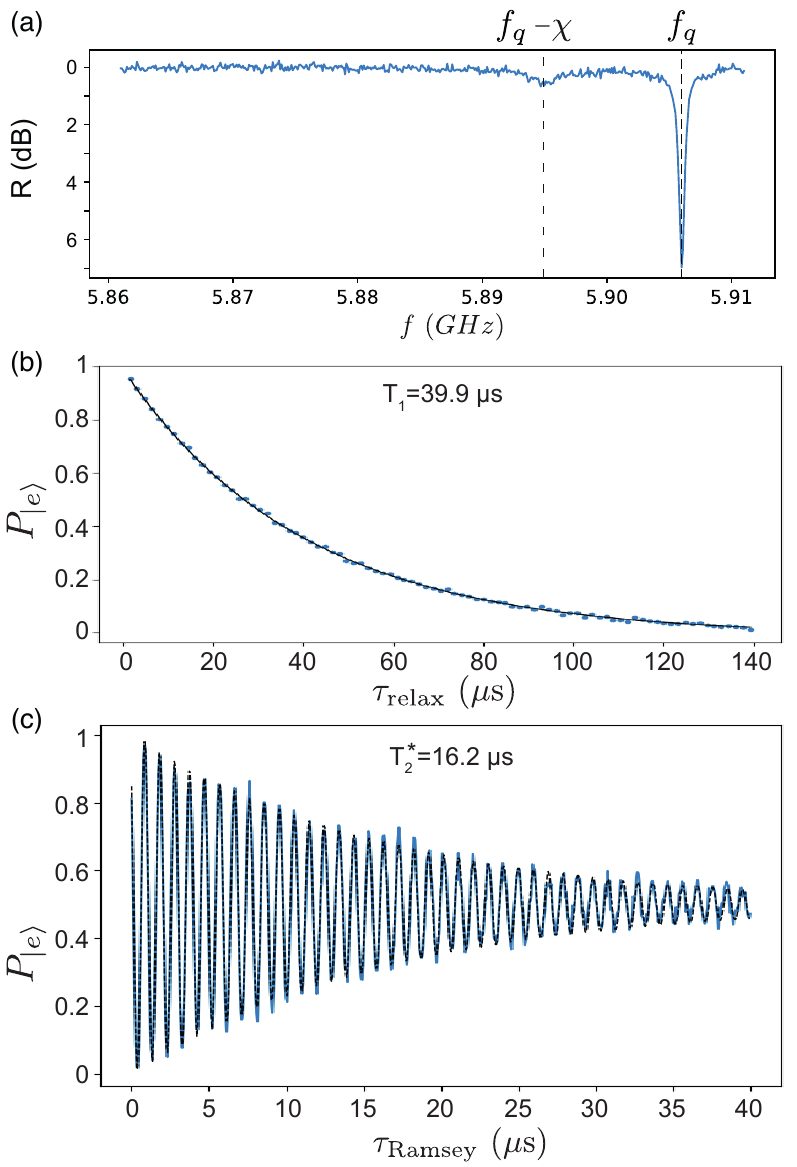}
    \caption{\label{fig7}
    Spectral and time-domain measurements of the transmon qubit. 
    (a) Two-tone spectroscopy of the transmon qubit. Here the excitation tone is provided by a commercial microwave source.
    (b) $T_1$ relaxation curve: qubit population as a function of delay time $\tau_{\mathrm{relax}}$ after a $\pi$ pulse, which was calibrated from the Rabi oscillations. The solid line is a fit to an exponential decay that yields $T_1 \approx \SI{39.9}{\us}$.  
    (c) Ramsey oscillations: qubit population as a function of delay time between two $\pi/2$ pulses $\tau_{\mathrm{Ramsey}}$ detuned at $f_d-f_q \approx \SI{1}{MHz}$. The fitting (solid line) yields $T_2^* \approx \SI{16.2}{\us}$.
    }
\end{figure}

The complete design of the circuit investigated in Section~\ref{sectiontransmon}, which mostly consists in a transmon qubit capacitively coupled to a microwave resonator, is described in Refs.~\cite{Balembois2023,Balembois2024}. We characterized this circuit-QED architecture using standard microwave instruments. 
Fig.~\ref{fig7}a shows the transmon qubit spectrum. The extracted transmon frequency $f_{q} \approx \SI{5.9060}{GHz}$ matches the value obtained from two-tone Josephson spectroscopy. Note however that the measured linewidth is here about 1 order of magnitude smaller than the one obtained via two-tone Josephson spectroscopy, the latter being limited by the radiation linewidth of the Josephson emitter. Finally, we also observe a second transition at $f_q-\chi$ associated to the finite population of Fock state $|n=1\rangle$ in the cavity.

Going further, we have probed the qubit quantum coherence by performing operations in the time domain. The strong coupling between qubit and cavity allows one to perform single-shot measurements of the qubit state. We could thus measure the probability $P_{|e\rangle}$  of the qubit to be in the excited state as a function of time. Using Rabi oscillations, one can calibrate $\pi$ and $\pi/2$-pulses. Thus, by applying a $\pi$-pulse that prepares the qubit in state $|e\rangle$, we could measure after a variable delay an exponential decay trace with an energy relaxation time $T_1 \approx \SI{39.9}{\us}$ (Fig.~\ref{fig7}a). Finally, we have measured the dephasing for an equal qubit superposition $(|g\rangle+|e\rangle)/\sqrt{2}$, by applying two $\pi/2$-pulses detuned at $f_d-f_q \approx \SI{1}{MHz}$ and separated by a varying delay time. Fig.~\ref{fig7}b shows the corresponding Ramsey oscillations, which indicate coherent precession of the qubit state in the Bloch sphere around the z-axis. From the decay, we could extract a dephasing time $T_2^{*} \approx \SI{16.2}{\us}$. This value being much smaller than $2T_1$, our qubit coherence is dominated by pure dephasing with a characteristic time $T_\phi \approx \SI{20}{\us}$.

% \begin{figure}
%     \centering
%     \includegraphics[width=\linewidth]{Fig8.pdf}
%     \caption{\label{fig8}
%     Time-domain measurements of the transmon qubit. 
%     (a) $T_1$ relaxation curve: qubit population as a function of delay time $\tau_{\mathrm{relax}}$ after a $\pi$-pulse, which was calibrated from the Rabi oscillations. The solid line is a fit to an exponential decay that yields $T_1 \approx 39.9~\mu$s.  
%     (b) Ramsey oscillations: qubit population as a function of delay time between two $\pi/2$-pulses $\tau_{\mathrm{Ramsey}}$ detuned at $f_d-f_q \approx 1$ MHz. The fitting (solid line) yields $T_2^* \approx 16.2~\mu$s.}
% \end{figure}

\section{Coupling schemes}
\label{Coupling}

One key aspect of two-tone Josephson spectroscopy is to probe high-frequency systems while using a detector with a resonant frequency $f_r$ that can easily be accessed with standard instruments and components ($<\SI{20}{GHz}$). One thus needs to engineer the proper coupling that will result in a measurable frequency shift $\chi$ of the detector. In section \ref{sectiontransmon}, the coupling between the transmon qubit and the detector is in the dispersive regime, with a detuning larger than the coupling strength yet small enough to allow the exchange of virtual photons. Following Ref.~\cite{koch_charge-insensitive_2007}, the frequency shift reads $\chi=-2 E_c ~g^2 / (\Delta (\Delta-E_c))$, with $E_c$ the qubit charging energy, $g$ the coupling strength and $\Delta =|f_r-f_q|$ the detuning. Although increasing the transmon qubit frequency would be a whole new work in itself, one can estimate the dispersive coupling in the regime of very large detuning. Considering, for instance, an SNS-based transmon qubit resonating at 100~GHz (designed with $E_J=\SI{500}{GHz}$, $E_C=\SI{2.4}{GHz}$, $g=\SI{250}{MHz}$ and $f_r=\SI{16}{GHz}$), we estimate $\chi \approx \SI{44}{kHz}$, which would translate into a large-enough signal to measure using a high-Q resonator.

More crucially, two-tone Josephson spectroscopy is more adapted to coupling in the adiabatic regime, where the detuning between the system and the detector is so large that virtual processes are strongly suppressed. Still, exciting the system can result in a measurable frequency shift of the detector, which does not depend on the detuning. The adiabatic regime indeed accounts for the renormalization of the resonator frequency by the effective capacitance or inductance of the system. For instance, in the case of a Josephson weak link hosting Andreev bound states, one should achieve large couplings by exploiting the variation of the Josephson inductance in the adiabatic regime, as described in Refs.~\cite{Park2020, Metzger2022}. Using typical parameters, we estimate a frequency shift of several 100~kHz when exciting the Andreev transition.

\section{Complementary two-tone Josephson spectroscopy measurements}
\label{Complementary}

\begin{figure}
    \centering
    \includegraphics[width=\linewidth]{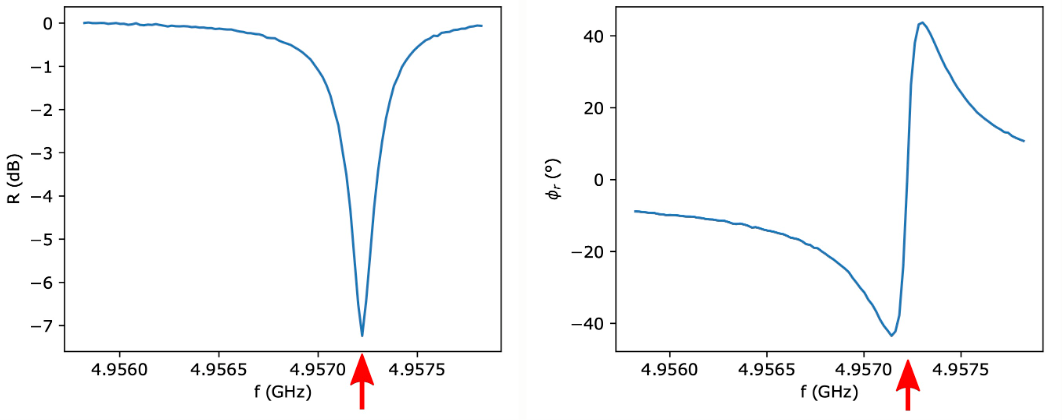}
    \caption{\label{fig9bis}
    One-tone spectroscopy of the $\lambda/4$ resonator used in Section~\ref{sectionlambda4} measured either via the magnitude (left) or the phase (right) of the reflected signal.}
\end{figure}

Fig.~\ref{fig9bis} shows the one-tone spectroscopy of the $\lambda/4$ resonator used in Sec.~\ref{sectionlambda4}, measured at low power and around 5~GHz. This measurement allows us to extract the fundamental frequency $f_0=\SI{4.9572}{GHz}$. The two-tone spectroscopy measurements are performed by probing the cavity response close to its resonance frequency (as indicated by the red arrow), which translates into a large phase response when a higher-frequency mode is excited.

\begin{figure}
    \centering
    \includegraphics[width=0.85\linewidth]{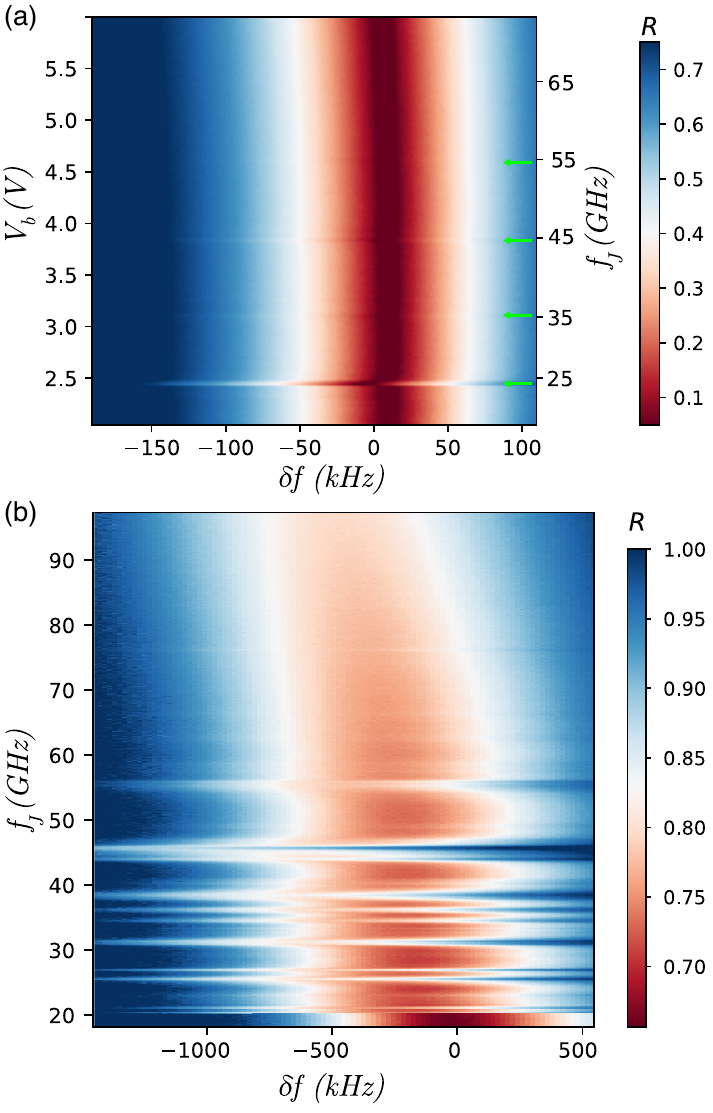}
    \caption{\label{fig9}
    (a) Two-tone Josephson spectroscopy for a sample similar to the one of Fig.~\ref{fig3}, with a smaller coupling between the resonator and the superconducting loop. Reflection coefficient as a function of the bias voltage $V_b$. The frequency $\delta f$ is referred to the resonance frequency of the cavity in the absence of bias voltage. The green arrows indicate the modes $m=2$, $3$, $4$, and $5$ that we detect in this experiment.
    (b) Two-tone Josephson spectroscopy for a sample with a much larger emitter-detector coupling. In this sample, we detect many more modes than the harmonic resonances of the cavity, which we attribute to hybridization of the resonator and the emitter junction circuit lines.}
\end{figure}

In Fig.~\ref{fig9}a, we present two-tone Josephson spectroscopy measurements performed in an additional sample similar to the one of Fig.~\ref{fig3}. The mutual inductance between the resonator and the superconducting loop was however designed to be smaller, which reduces the Kerr effect. A dip in the amplitude of the reflected signal indicates the position of the $m=0$ mode of the microwave resonator. As we sweep the bias voltage, the Josephson frequency happens to match the frequency of a higher mode. As the latter is populated by microwave photons, the $m=0$ resonance shifts by a few kHz, which allows the detection of high-frequency modes.

In Fig.~\ref{fig9}b, we present measurements performed on another device, where the coupling between the Josephson emitter and the resonator was made much stronger. We detect many more resonances than in the measurements presented in the main text; however, they do not correspond to the bare resonances $f_m=(2m+1)f_0$ of the $\lambda/4$ resonator. We believe that in this case the spectroscopy is too invasive, because the modes of the resonator are perturbed by the emitter due to the strong coupling.

\section{Two-tone Josephson spectroscopy with a tunnel Josephson junction}
\label{tunnel}

\begin{figure}
    \centering
    \includegraphics[width=\linewidth]{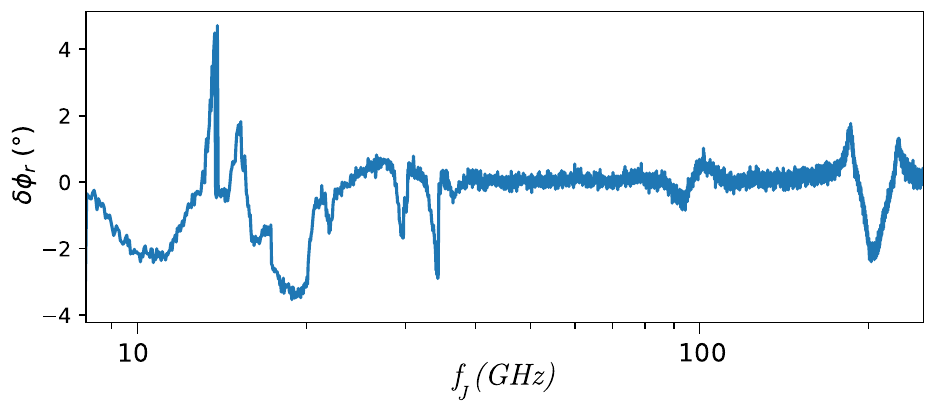}
    \caption{\label{fig10}
    Two-tone Josephson spectroscopy performed using a tunnel Josephson junction as an emitter and a 7~GHz resonator as a detector.  
    We interpret the features at 100 and 200~GHz as excitation of quasiparticles.
    }
\end{figure}

One of the reasons we opt for an SNS junction over an SIS tunnel junction as the Josephson emitter is due to the SNS junction's relatively constant impedance when a voltage is applied across its terminals, compared to the tunnel junction. In Fig.~\ref{fig10}, we show measurements of a two-tone Josephson spectroscopy, similar to Fig.~\ref{fig3}d, but conducted with an SIS junction as the emitter. We observed significant signal variations with numerous structures resembling resonances. However, they do not necessarily signify mode detections but are sometimes simply due to changes of the resonance frequency of the resonator fundamental mode ($m=0$) caused by impedance variations of the SIS junction. Consequently, conducting two-tone Josephson spectroscopy with this type of junction is challenging.

\section{Sign of the resonator frequency shift}
\label{theory}

In Fig. \ref{fig3}e of the main text, we observe that the dispersive coupling $\chi_{0,6}$ between modes $m=0$ and $m=6$ changes sign depending on whether the normalized flux in the superconducting loop, $\phi$, is $0$ or $\pi$. This behavior is actually consistent with the expected electromagnetic modes of the $\lambda/4$ resonator.

We can understand this by modeling our device as a resonator of length $l$ and characteristic impedance $Z_0$, which is open at one end and terminated by an inductance $L$ at the other, as shown in Fig. \ref{fig3}a. The fundamental mode then has a resonance frequency given approximately by
$f_0 \approx f_{\lambda/4} \left(1 - \frac{L}{L_r}\right)$,
where $L_r = \frac{Z_0}{4f_0} \gg L$ is the total inductance of the resonator, and $f_{\lambda/4} = \frac{v}{4l}$ is the bare resonance frequency, with $v$ being the propagation speed of microwave photons. The contribution of the superconducting loop to the inductance $L$ is expressed as $\frac{-M^2}{L_J + L_g}$, where $M$ is the mutual inductance between the two, $L_g$ is the geometrical inductance of the loop, and $L_J$ is the Josephson inductance of the SNS junction. A slight change in Josephson inductance $\delta L_J$ results in a frequency shift
$\delta f_0 \approx -\left(\frac{M}{L_g+L_J}\right)^2 \frac{\delta L_J}{L_r} f_0$,
which can be either positive or negative, depending on the sign of $\delta L_J$.

Calculating the change in Josephson inductance, $\delta L_J$, due to mode population requires knowledge of the exact current-phase relation (CPR) of the junction. To estimate this change, we use a simplified model where the SNS junction has a sinusoidal CPR, similar to that of a tunnel junction: $I(\phi) = I_c \sin{\phi}$. When a high-frequency current, $I_{ac}$, flows through the junction, its critical current is renormalized as $I_c(I_{ac}) = I_c J_0(I_{ac}/I_c)$, where $J_0$ is a Bessel function of the first kind. The Josephson inductance then becomes
$L_J(I_{ac}) = \frac{\hbar}{2e} \frac{1}{I_c(I_{ac}) \cos{\phi}}$.
In the limit of small excitation current ($I_{ac}\ll I_c$), we find
$\delta L \approx L_J(\phi) \left( \frac{I_{ac}}{2I_c} \right)^2$. Thus, the sign of the inductance change $\delta L$ is determined by the sign of $\cos{\phi}$. Near $\phi = 0$, $\delta L$ is positive, meaning the resonator frequency decreases as $I_{ac}$ increases (\emph{i.e.} $\delta f<0$). Conversely, near $\phi = \pi$, the resonator frequency increases at finite $I_{ac}$ (\emph{i.e.} $\delta f>0$).

Finally, knowing that adding a single photon to the mode at frequency $f_m$ is equivalent to setting $I_{ac} = \sqrt{\frac{hf_m}{L_r}} \frac{M}{L_g+L_J(\phi)}$, the dispersive shift between modes $m$ and $0$ can be written as
$\chi_{0,m} \approx \left(\frac{M}{L_g+L_J}\right)^4 \frac{L_J}{L_r} \frac{h f_m}{L_r I_c^2} f_0$.
In our experiment, we estimate $L_r\approx \SI{2.5}{nH}$, $M\approx 5-\SI{10}{pH}$, and $L_g\approx \SI{30}{pH}$. The SNS junction was designed with a critical current of $I_c\approx \SI{2}{\uA}$, corresponding to a Josephson inductance of $L_J \approx \SI{160}{pH}$. This critical current is approximately 10 times smaller than that of the emitter junction, due to the normal metal part being twice as long ($L = \SI{600}{nm}$), and $I_c$ scaling as $L^{-3}$ for SNS junctions in the long-junction limit. The dispersive shift $\chi_{0,6}$ is thus on the order of $1-10$~Hz in our system, with its sign dependent on the sign of $L_J(\phi)$. Based on the measured phase shift, this indicates that our Josephson emitter populates the mode with approximately thousands of photons.

\section{Bandwidth of the Josephson emission in the high-frequency modes spectroscopy experiment}
\label{Linewidth}

\begin{figure}
    \centering
    \includegraphics[width=\linewidth]{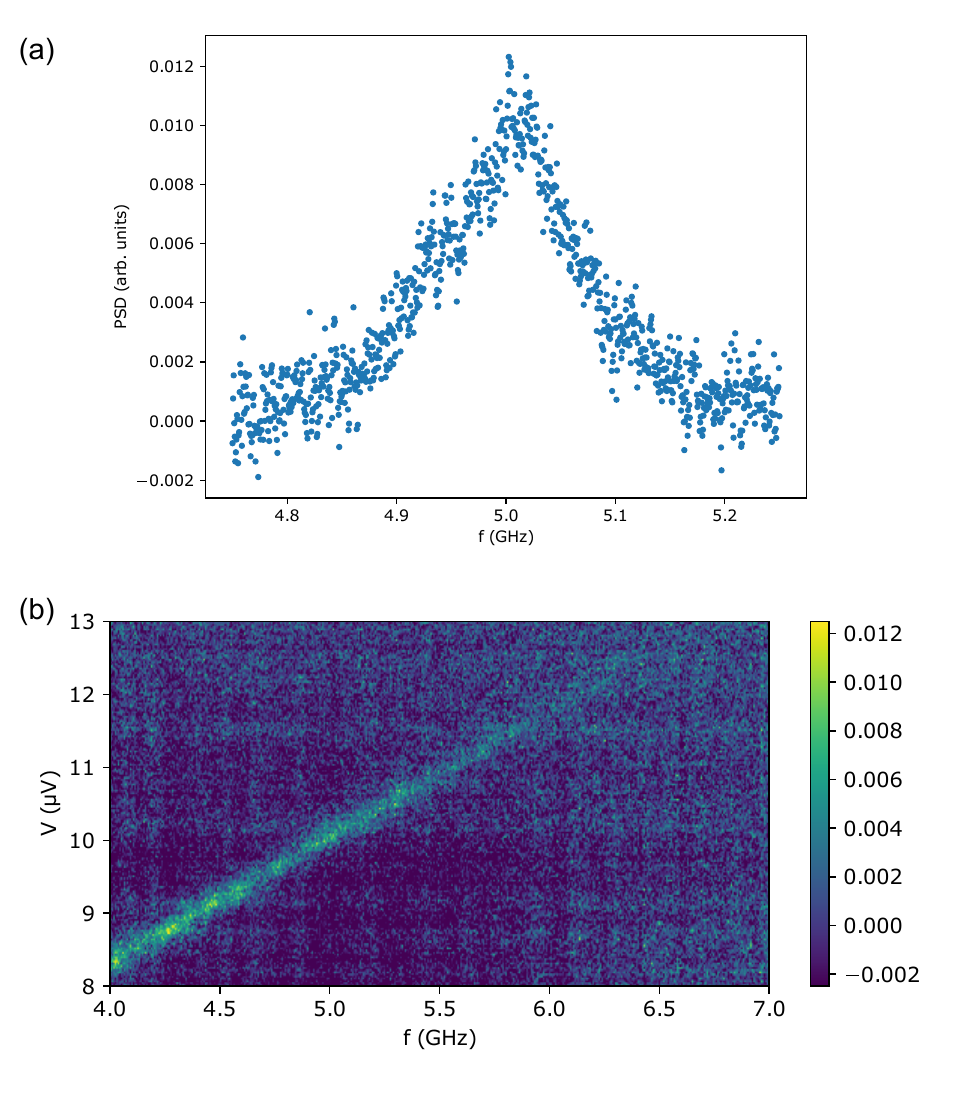}
    \caption{\label{fig11}
   Josephson power $P$ measured without the $4-8$~GHz passband filter. (a) $P$ as a function of frequency $f$ for a voltage across the SNS emitter $V_J=\SI{10.3}{\uV}$. (b) $P$ as a function of both $f$ and $V_J$. }
\end{figure}

In the experiment shown in Fig. \ref{fig3}, the peaks of the $\lambda/4$ modes that we observe have a width of approximately 500~MHz. This broadening is not related to the lifetime of these modes but rather to the linewidth of the Josephson emission, which is 100 times larger than in the other experiments presented here. This difference is due to the use of a different filtering setup, which, unlike the other experiments, did not include the $4-8$~GHz bandpass filter. We observed that the voltage noise across the junction was significantly higher, resulting in an emission linewidth that was increased by a factor of 100. This is illustrated in Fig.~\ref{fig11}, where the emission linewidth is several hundred MHz, caused by the increased voltage noise across the junction.

\end{appendix}
\bibliography{Bib_2TeJoS}

\end{document}